\documentclass[aps,prb,twocolumn,superscriptaddress,floatfix]{revtex4}
\usepackage{physics}
\usepackage{float}
\usepackage{graphicx}
\usepackage{color} 

\begin{document}
\title{Projected Development of COVID-19 in Louisiana}% Prior to Mitigation Efforts}
 
\author{Ka-Ming Tam}
\affiliation{Department of Physics \& Astronomy, Louisiana State University, Baton Rouge, Louisiana 70803, USA}
\affiliation{Center for Computation \& Technology, Louisiana State University, Baton Rouge, Louisiana 70803, USA}

\author{Nicholas Walker}
\affiliation{Department of Physics \& Astronomy, Louisiana State University, Baton Rouge, Louisiana 70803, USA}

\author{Juana Moreno}
\affiliation{Department of Physics \& Astronomy, Louisiana State University, Baton Rouge, Louisiana 70803, USA}
\affiliation{Center for Computation \& Technology, Louisiana State University, Baton Rouge, Louisiana 70803, USA}

\date{\today}

\begin{abstract}

At the time of writing, Louisiana has the third highest COVID-19 infection per capita in the United States. The state government issued a stay-at-home order effective March 23rd. We analyze the projected spread of COVID-19 in Louisiana without including the effects of the  stay-at-home order. We predict that a large fraction of the state population would be infected without the mitigation efforts, and would certainly overwhelm the capacity of Louisiana health care system. We further predict the outcomes with different degrees of reduction in the infection rate. More than 70\% of reduction is required to cap the number of infected to under one million. 
\end{abstract}

\maketitle 
\section{Introduction}

The identification and verification of human-to-human transmission of the coronavirus  disease
(COVID-19) in early January of 2020 in Wuhan, China triggered the start of  a worldwide pandemic.
As of April 5, there are more than 1.2 million confirmed cases and more than 67,000 deaths attributed to COVID-19. The first case in the US was confirmed in Washington State on January 20. The number 
of reported cases until early March was rather low.
%. While the mandatory quarantine for anyone who had traveled to China has undoubtedly slowed the development in the USA, t
The exceedingly slow spreading rate in these early months may be partially due to the lack of adequate testing, which remains a major issue at the time of writing.
%By late February, various clusters of community spreading were identified in Italy. 
The cases dramatically increased in the USA in early March, 
%resulting in the enforcement of a federal travel ban on individuals that visited any of 28 Europe countries within the prior 14 days. 
with most cases in the states of Washington, New York, and California. It was not until March 9 that the first case in Louisiana was identified. 

The growth rate of infections in Louisiana has been alarming since the confirmation of the first case. Louisiana state government responded swiftly by closing all K-12 public schools on March 13. On March 16, public gatherings of more than 50 people were prohibited, and bars, bowling alleys, casinos, fitness facilities, and movie theaters were closed. Furthermore, a stay-at-home order was issued 
on March 23.

%The testing capacity is delayed and limited in Louisiana, a concern that affects the rest of the country as well. Even though the situation has been reported to have improved by the end of March, a
Adequate testing for COVID-19 remains limited in the USA. 
%resource that is not easily accessible to the public. 
For this reason, accurately predicting the trajectory of the spread of COVID-19 by relying on the number of confirmed cases alone is a rather questionable approach. While the Susceptible-Infected-Recovered (SIR) model may well describe the dynamics of the spreading \cite{Huppert,Kermack_McKendrick}, accurate predictions rely on knowing the number of confirmed cases, which is severely hampered by the limitations of testing. This is particular significant in the 
early stages of the spread of the disease when the percentage of people tested is very small, and
the spread by infected people who are asymptomatic is very significant.

%The limited testing facilities for COVID-19 lead to a problematic situation in which most of the infected persons who are asymptomatic are not tested. This exacerbates the spread of the virus through undetected infected persons. 
Alternatively, the number of fatalities attributed to COVID-19 may be a more reliable parameter 
for tracing the dynamics of the virus spread. 
%Most of the severely ill patients with symptoms of COVID-19 have presumably been tested and identified. 
Combining this information with the mortality rate can be a better strategy to predict the number of cases than relying on the confirmed infection count alone.
%, as the casualty count is likely to be less affected by inadequate testing. 
The goal of this paper is to extract the dynamics of COVID-19 in Louisiana from the data of the death count supplemented with the confirmed cases. %casualties, and to predict the virus spread in the absence of mitigation efforts. 
%, particularly before the stay-at-home order initiated on March 23.
We then run several scenarios with different reduction of the infection rate and calculate the number of people 
infected in each case. We conclude with suggestions to improve the model and, as consequence, its predictions.

\section{Model} \label{Sec:Model}
Our model is based on the Susceptible-Infected-Recovered (SIR) model\cite{Crokidakis,Bin,Pedersen,Calafiore,Bastos,Gaeta1,Gaeta2,Vrugt,Schulz,Zhang,Amaro,DellAnna,Sonnino,Notari,Simha,Acioli,Zullo,Sameni,Radulescu,Roques,Teles,Piccolomini,Brugnano,Giordano,Zlatic,Baker,Biswas,Zhang_Wang_Wang,Chen,Lloyd} with the modification  of including the number of quarantined people (Q), as has been considered elsewhere. \cite{Crokidakis,Bin,Pedersen} The equations defining the model are the following:

\begin{equation}
\label{Eq:Seq} 
\dv{S(t)}{t} = -\beta \frac{S(t)I(t)}{N},
\end{equation}

\begin{equation}
\label{Eq:Ieq} 
\dv{I(t)}{t} = \beta \frac{S(t)I(t)}{N} -\qty(\alpha +\eta)I(t),
\end{equation}

\begin{equation}
\label{Eq:Qeq} 
\dv{Q(t)}{t} = \eta I(t)-\gamma Q(t),
\end{equation}

\begin{equation}
\label{Eq:Req} 
\dv{R(t)}{t} = \gamma Q(t) +\alpha I(t),
\end{equation}
where $N$ is the total population size, $S$ is the susceptible population count, $I$ is the unidentified infected population count, $Q$ is the number of identified cases, and  $R$ includes the number of recovered and dead patients.
The model is characterized by the following parameters: $\beta$ is the infection rate, $\eta$ is the detection rate, $\alpha$ is the recovery rate of asymptomatic people, and $\gamma$ includes the recovery rate and the casualty rate of the quarantined patients. This model is equivalent to the standard SIR model if we are not interested into differentiating 
between Q and R. 

We further assume that the rate of increase in the number of casualties is proportional to the number of infected at the early stage of the epidemic,
\begin{equation}
\label{Eq:Ceq} 
\frac{dC(t)}{dt} = \delta I(t),
\end{equation}
where $\delta$ is the mortality rate.
This is a good approximation at the beginning of the virus spread when the number of quarantined patients is a small
percentage of the total population. This equation is not combined in any way with Eq. 1-4, it is only used to estimate 
the model parameters at the start of the epidemic.

\section{Analysis} \label{Sec:Analysis}

We first consider Eq. \ref{Eq:Ieq}, assuming the susceptible population count is very close to that of the total population, $S \sim N$, which is justifiable at the beginning of the epidemic since only a small fraction of the population is infected. With this assumption one can decouple the infected population count from the other parameters to obtain: \cite{Crokidakis,Pedersen}
\begin{equation}
\label{Eq:I(t)}
I(t) = I(0)\exp\qty[\qty(\beta -\qty(\alpha +\eta))t].
\end{equation}

Solving Eq. \ref{Eq:Ceq},  the casualty count as a function of time can be written as 
\begin{equation}
\label{Eq:C(t)}
C(t) = \frac{\delta I(0)}{\beta-(\alpha+\eta)}\exp\qty[\qty(\beta- \qty(\alpha+ \eta))t].
\end{equation}

The exponential growth of the number of fatalities at the beginning of the epidemic should represent the spreading of COVID-19 reasonably well since the mechanisms for slowing the dynamics, such as improved detection and social distancing, are delayed in time

\begin{figure}[t]
\centerline{\includegraphics[width=0.5\textwidth]{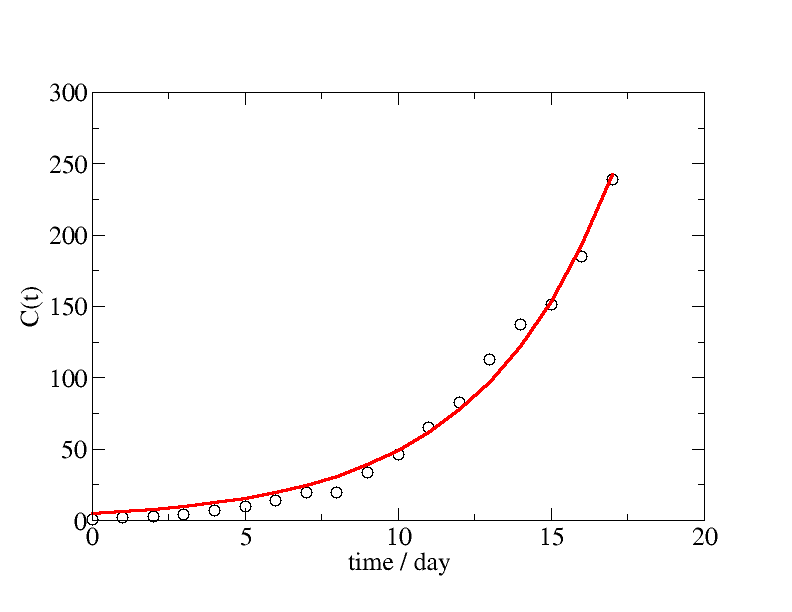}}
\caption{COVID-19 casualties in Louisiana, $C(t)$, as a function of time with March 14 as day 0, data is represented by circles.
The parameters of Eq. \ref{Eq:C(t)} are fit to the data, providing an approximation to the number of deaths as a function of time: $C(t) \approx 5.04 \exp\qty[0.228t]$.}
\label{Fig:C(t)}
\end{figure}

By fitting the available fatalities data (see Appendix) between %the first recorded death on
March 14 and 31 to Eq. \ref{Eq:C(t)}, the parameters of the model can be determined.
%The expression $\qty(\beta- \qty(\alpha+ \eta))$ is estimated via fitting Eq. \ref{Eq:C(t)} to the available casualty data provided in the appendix.  
Fig. \ref{Fig:C(t)} displays the fit which provides an estimate of $C(t) \approx 5.04 \exp\qty[0.228t]$. The dynamics (exponent) is thus given as $\beta-(\alpha+\eta) = 0.228$. 
From the value of the exponent we can estimate the time for doubling the casualties count: 
$ln(2)/0.228 \approx 3.04$ days.
Moreover, the proportionality constant can be used to estimate the initial number of infections $I(0)$ if the mortality rate $\delta$ is known.

The mortality rate is estimated by combining the accumulated mortality rate data and the median time between infection and death. It is estimated that the median time between infection and the onset of symptoms is about five days, while the 
median time between the onset of symptoms and death is eight days.\cite{WHO,Anderson,Li,Linton} It is worth noting that 
the distribution of these time periods is close to a log-normal, thus a more sophisticated analysis should include the effects of the non-self-averaging behavior of the distribution. Only the median values are used in the present work.

The accumulated mortality rate is estimated to be 2.3\% \cite{Wu}. Notably, the mortality rate does indeed vary by region. This may be due to the rate of testing as well as the capacity of health care facilities. For areas in which health care facilities have been overrun, the death rate would be much higher. Notwithstanding these uncertainties, assuming that the health care facilities have not yet been overrun, the mortality rate is estimated to be $\displaystyle \delta \approx \frac{0.023}{5+8} \approx 0.0018$. This also provides an estimation of the number of persons who carries the virus but not detected at day 0, $I(0)$, which is given as $\displaystyle I(0) \approx \frac{5.04}{\delta} \cross 0.228 \approx 650$. This reveals that even as early as March 14, the number of infected people is already at the order of hundreds. %{\bf CHECK THIS Sentence: We emphasize that this number include all confirmed and unidentified cases.}

Now we consider the number of confirmed cases at the start of the epidemic, $P(t)$. This is given by the 
sum of $Q(t)$ and $R(t)$ subtracted by the number of persons who recovered without being tested. The rate of 
change on the number of reported cases can be obtained by combining Eqs. \ref{Eq:Qeq} and \ref{Eq:Req} 
and subtracting $\alpha I(t)$:  
\begin{equation}
    \frac{dP(t)}{dt} = \eta I(t)
    \label{Eq:P(t)}
\end{equation}
with $I(t)$ given by Eq. \ref{Eq:I(t)}, we obtain:~\cite{Crokidakis,Pedersen}
\begin{equation}
P(t) \approx \frac{\eta}{\beta -\alpha -\eta}I(0) \exp\qty[0.228t].
\end{equation}

\begin{figure}
\centerline{\includegraphics[width=0.5\textwidth]{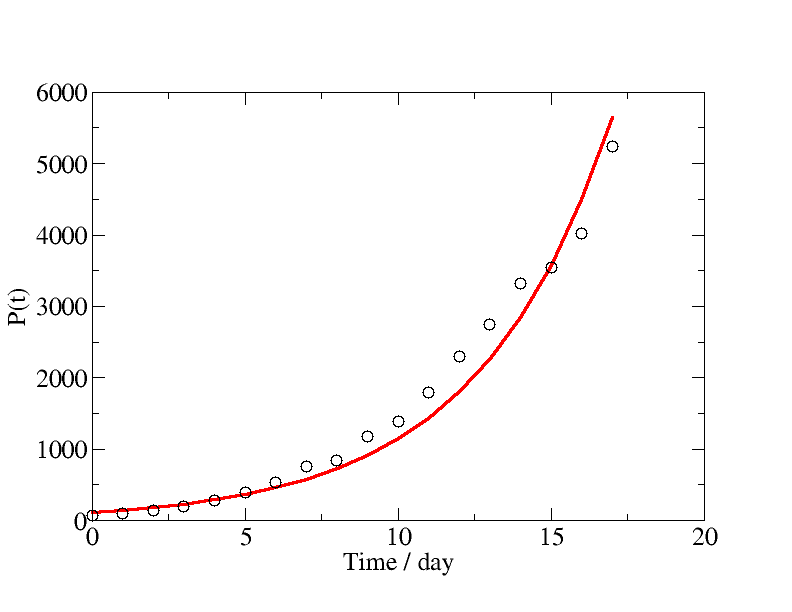}}
\caption{The number of confirmed cases of COVID-19 in Louisiana, $P(t)$, as a function of time 
with March 14 as day 0. 
The data is represented by circles, the red line is the best fit,
$P(t) \approx 117 \exp\qty[0.228t]$. Note, we only fit the coefficient in front, the exponent is given by from the death count, see fig. \ref{Fig:C(t)}.}
\label{Fig:I(t)_future}
\end{figure}

By fitting the number of confirmed cases (Fig.\ref{Fig:I(t)_future}) we find $P(t) \approx 117 \exp\qty[0.228t]$, which provides the estimate $\displaystyle \frac{\eta}{\beta-\alpha-\eta}I(0) \approx 117$. Since $I(0)$ and 
$(\beta -\alpha -\eta)$ are known from fitting to the number of casualties, we find $\eta \approx 117 \cross \frac{0.228}{650} \approx 0.041$. 
There remains one parameter to be determined, the recovery rate of asymptomatic people, $\alpha$. Assuming that the average time or recovery or dead are both $13$ days, and half of the infected never show any symptoms thus they are not been tested \cite{Mizumoto}. We can estimate $\alpha=0.5/13 \approx 0.0385$. This is probably the upper bound of the estimate, in reality this could be smaller. This additionally provides the value for 
$\beta$ as $0.307$.
%Fig. \ref{Fig:I(t)_future} shows the data for the number of person who have been infected for the fit with the exponent of $0.23$.

%By the discussed fitting procedure we found all the parameters relevant for the model: $I(0) = 640$, $\alpha+\eta = 0.041$, and $\beta = 0.269$. 

\begin{figure}
\centerline{\includegraphics[width=0.5\textwidth]{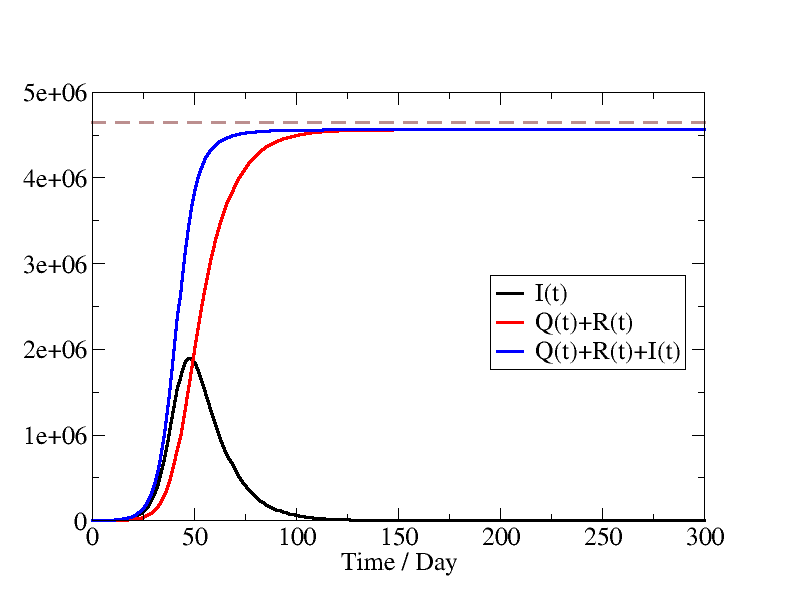}}
\caption{The number of unidentified but infected people, $I(t)$, the number of quarantined and recovered people,  Q(t)+R(t), and the number of people who are ever infected, Q(t)+R(t)+I(t), as a function 
of time since March 14. Horizontal dashed line at the total population of Louisiana, 4.65 million.}
\label{Fig:I(t)}
\end{figure}

With these parameters, Eqs. 1-4 can be solved and used to predict the spread of the disease. 
Fig. \ref{Fig:I(t)} displays the time evolution of the number of unidentified persons who carry the virus, $I(t)$, the number of persons who are either in quarantine or recovered, $Q(t)+R(t)$, and the total number of persons 
who have ever been infected, $Q(t)+R(t)+I(t)$.  
The number of infections but unidentified, $I(t)$, grows exponentially, as expected from Eq. \ref{Eq:I(t)}, 
at the initial stage, and this behavior continues until about day 25, when around 100,000 people are infectious. The exponent of $\sim 0.228$ suggests the number of cases double approximately every three days, which seems to be consistent with the data in many areas of the world 
before the mitigation efforts are kicked in. 
After day 25, the rate of increase slow down due to the combination of the decrease 
on the number of susceptible (uninfected) people and the increase on the number of recoveries. 
The number of infected cases ceases to grow exponentially, but rather becomes a stable but constant 
increases until peaking at around day 50, corresponding to early May. On the other hand, the number of quarantined and  recovered people resembles a logistic function.
%More than 80\% of the population may be actively carrying the virus at this time. 

%Perhaps more alarmingly, the number of carriers will remain high for a long period of time. More than $20\%$ of the population will still carry the virus after more than half a year. Essentially, it diminishes slowly  only after a large portion of the population are infected.

To compare with other states which already have widespread epidemic, we use the described method to calculate the infection rate ($\beta$), the testing rate ($\eta$), and the reproduction number ($R_{0}=\beta/(\eta+\alpha$)) of selected states. Result are displayed in Table \ref{Table:exponents}. Note than  the reproduction number of Louisiana 
is the highest among the states listed in the table. %{\bf This number .....}

\begin{center}
\begin{table}
 \begin{tabular}{||l c c c||} 
 \hline
 State & $\beta-\alpha-\eta$ & $\eta$ & $R_{0}$ \\ [0.5ex] 
\hline\hline
Louisiana & 0.228 & 0.041 & 3.87\\
Florida & 0.198 & 0.114 & 2.30\\
Georgia & 0.161 & 0.054 & 3.74\\
Texas & 0.206 & 0.114 & 2.35 \\
California & 0.205 & 0.083 & 3.69 \\
Illinois & 0.237 & 0.085 & 2.92 \\
New Jersey & 0.287 & 0.079 & 3.44 \\
New York & 0.231 & 0.070 & 3.13 \\[1ex] 
 \hline 
\end{tabular}
\caption{\label{Table:exponents}Table of the value of $\beta-\alpha-\eta$, $\eta$, and $R_{0}$ for different states. Note that the reproduction number, $R_0$, is in line with the other studies \cite{Imperial}. }

\end{table}
\end{center}

\begin{figure}
\centerline{\includegraphics[width=0.5\textwidth]{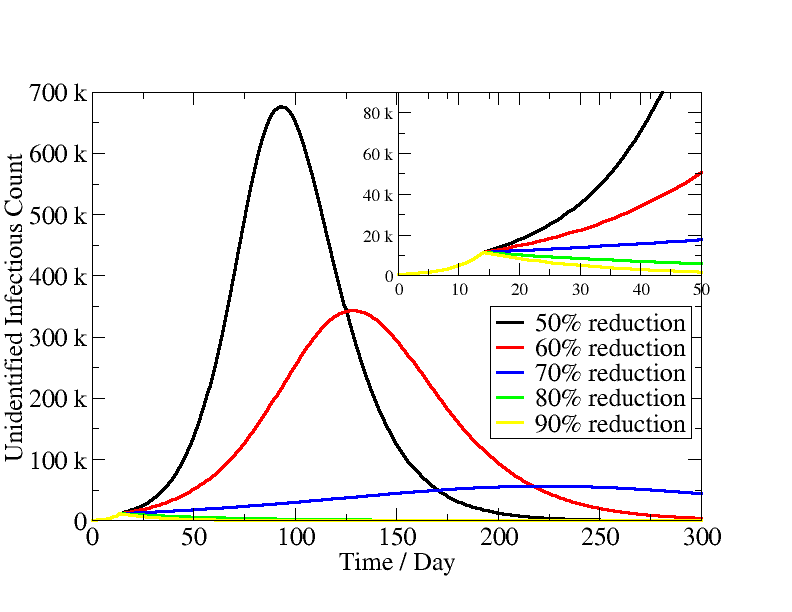}}
\caption{The number of people who are infected and carrying the virus without being identified, $I(t)$, as a function of time, with March 14 as day 0. We assume the mitigation efforts reduce the infection rate by 50 \%, 60\%, 70\%, 80\%, and 90\% from day 15 (6 days after the stay-at-home order), and the sum of the testing rate and recovery rate of asymptomatic people remains unchanged. The inset is a zoom for the first 50 days.}
\label{Fig:red_i}
\end{figure}

\begin{figure}[H]
\centerline{\includegraphics[width=0.5\textwidth]{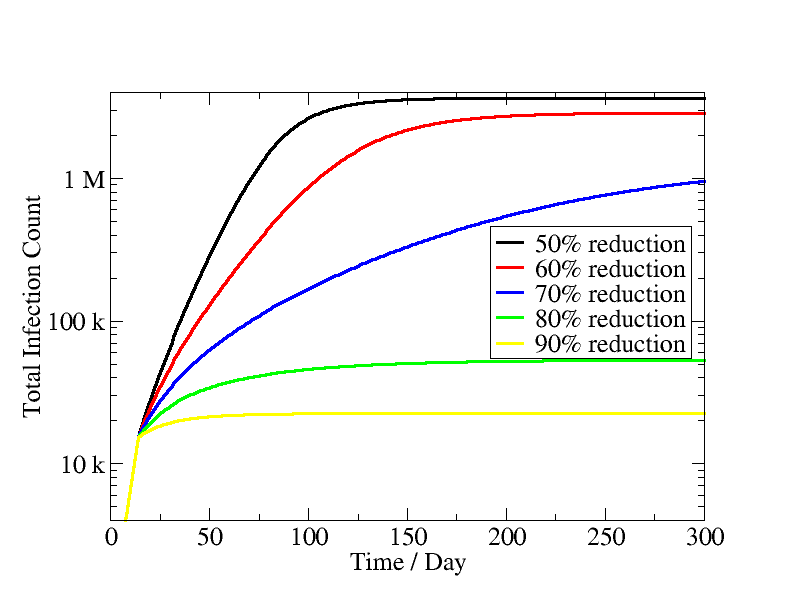}}
\caption{The number of people who are ever infected the virus (log scale), $I(t)+Q(t)+R(t)$, as a function of time, with March 14 as day 0. We assume the mitigation efforts reduce the infection rate by 50 \%, 60\%, 70\%, 80\%, and 90\% from day 15, and the sum of the testing rate and recovery rate of asymptomatic people remains unchanged.}
\label{Fig:red_iqr}
\end{figure}

Within the present model, there are two major routes to slow the initial exponential growth of the epidemic, 
which is characterized  by the parameter $\beta -(\alpha + \eta)$. The first one is to decrease the infection rate, $\beta$. The second route is to increase the testing rate, $\eta$. To increase the recovery rate from unidentified persons, $\alpha$, can also reduce the spread, but it is unlikely to be achieved. 
As the stay-at-home order was issued on March 23, it is expected that the infection rate should be drastically reduced. We simulate new scenarios with the assumption that social contact is reduced so that the infection rate decreases by 50\%,  60\%, 70\%, 80\%, and 90\% starting at day 15. The results are shown in Fig. \ref{Fig:red_i} and \ref{Fig:red_iqr}. 

We find that there is a substantial drop of the active virus carriers even with a 50\% reduction in the infection rate.  However, the number of people who will be infected still exceeds one million if the reduction in the infection rate is smaller than 70\%. This suggests the importance of strict measures in social distancing. Perhaps it also suggests the importance of wearing basic protective gear to further reduce the infection rate. 

\section{Discussion}

There are many uncertainties in this simplified model which can be improved over time as more data become available. Improvement can be achieved by including additional factors, such as correlation with different age groups, correlation with the health condition of the population, the availability of public health care, the effect of higher ambient temperature and humidity, and many others. Some of those factors are likely beyond the SIR model which implicitly assume that the population is homogeneous and well mixed, and that infection occurs without time delay. However, given the rather limited data available today, it is not clear that more sophisticated models may provide much better predictions. 

In spite of the rather simple model being employed in this analysis, it provides a baseline for the spread of the COVID-19 in Louisiana in the absence of mitigation efforts. The situation is clearly dire, as a very large fraction of the population will get infected with a peak on the number of infections around early May.

With the current mitigation efforts, we expect the infection rate will be greatly reduced. Currently, we do not have data to support the effectiveness of current mitigation efforts as the trend still fits rather well to the initial stage of exponential growth. 

%The effects may be appreciable after a week or two more, as the average onset of the symptoms is presumably about five days.
%In this paper we assume that the effect of the mitigation efforts can be captured with two assumptions: 1. Mitigation provides reduction in the infection rate. 2. The sum of the testing rate and the recovery rate from asymptomatic people can be increased. 

The main projection from this work is that more than 70\% of reduction in the infection rate is needed to keep the infected count below one million. Increasing testing capacity and providing protective gear to further reduce the infection rate seem to be reasonable measures. 

\section{Acknowledgment}
This work is funded by the NSF EPSCoR CIMM project under award OIA-1541079. This work used the high performance computational resources provided by the Louisiana Optical Network Initiative (http://www.loni.org) and HPC@LSU computing. Additional support (KMT) was provided by NSF Materials Theory grant DMR-1728457 and NSF Office of Advanced Cyberinfrastructure grant OAC-1931445.

%However, given the large number of people who have been infected and have not yet been identified, the time needed to slow the growth may be long.

\section{Appendix}

Number of people tested for COVID-19, people confirmed infected, and the resulting casualty count in Louisiana from March 14 to March 31 are shown in the Table \ref{Table:LA_data} \cite{LA_gov,LA_wiki,Advocate}

\begin{center}
\begin{table}
 \begin{tabular}{||c c c c||} 
 \hline
 Date & Tested & Confirmed & Death \\ [0.5ex] 
\hline\hline
Mar 14	&210	&77	&1\\
Mar 15	&247	&103	&2\\
Mar 16	&374	&136	&3\\
Mar 17	&531	&196	&4\\
Mar 18	&703	&280	&7\\
Mar 19	&899	&392	&10\\
Mar 20	&1,931	&537	&14\\
Mar 21	&3,302	&763	&20\\
Mar 22	&3,498	&837	&20\\
Mar 23	&5,948	&1,172	&34\\
Mar 24	&8,603	&1,388	&46\\
Mar 25	&11,451	&1,795	&65\\
Mar 26	&18,299	&2,305	&83\\
Mar 27	&21,359	&2,746	&119\\
Mar 28	&25,161	&3,315	&137\\
Mar 29	&27,871	&3,540	&151\\
Mar 30	&34,033	&4,025	&185 \\
Mar 31  &38,967 &5,237 &239\\[1ex] 
 \hline
\end{tabular}
\caption{\label{Table:LA_data} Number of test conducted, number of confirmed cases, and number of death from Mar 14 to Mar 31 from COVID-19 in Louisiana.}
\end{table}
\end{center}

%%%%%%%%%% Merge with supplemental materials %%%%%%%%%%
%\pagebreak
%\widetext
%\begin{center}
%\textbf{\large Supplemental Materials: Scaling behaviors at transition points}
%\end{center}
%%\twocolumngrid
%\setcounter{equation}{0}
%\setcounter{figure}{0}
%\setcounter{table}{0}
%\setcounter{page}{1}
%\makeatletter
%\renewcommand{\theequation}{S\arabic{equation}}
%\renewcommand{\thefigure}{S\arabic{figure}}
%\renewcommand{\bibnumfmt}[1]{[S#1]}
%\renewcommand{\citenumfont}[1]{S#1}

%\begin{figure}[!htb]
% \centerline{\includegraphics[width=0.8\textwidth]{Densityn1U22crossingotherend.eps}}
% \caption
%    {(Color online) 
%     Scaling behavior of the superfluid stiffness, $\rho_s$, for the integer density, $\rho=N/L=1$ and on-site interaction, $U=22t$. The $\rho_s L^2$ versus disorder strength, $\Delta$, for different system sizes cross at the critical disorder strength, $\Delta_c%%\approx 135t.$ The corresponding inverse temperatures for the linear system sizes are : $L = 6,  \beta t=6, L=8$, $\beta t=8$ and $L=16, \beta t=16$.
%The data points are based on simulation results for $100$ disorder realizations, the lines are guides to the eye.
%    }
%\label{Fig:Densityn1U22crossingotherend}
%\end{figure}

\end{document}